\documentclass[letter,10pt]{article}
\usepackage{makeidx}
\usepackage{graphicx}
\begin{document}
\title{\textbf{Coating thermal noise of a finite-size cylindrical mirror}}
\author{Kentaro Somiya$^\mathrm{a}$ and Kazuhiro Yamamoto$^\mathrm{b}$\\
\small{$^\mathrm{a}$\it{Theoretical Astrophysics, California Institute of Technology, Pasadena, California, 91125}}\\
\small{$^\mathrm{b}$\it{Max-Planck-Institut f\"ur Gravitationsphysik (Albert-Einstein-Institut), Callinstr. 38, 30167 Hannover, Germany}}}
\date{}

\twocolumn[
\maketitle
{\footnotesize Thermal noise of a mirror is one of the limiting noise sources in the high precision measurement such as gravitational-wave detection, and the modeling of thermal noise has been developed and refined over a decade. In this paper, we present a derivation of coating thermal noise of a finite-size cylindrical mirror based on the fluctuation-dissipation theorem. The result agrees to a previous result with an infinite-size mirror in the limit of large thickness, and also agrees to an independent result based on the mode expansion with a thin-mirror approximation. Our study will play an important role not only to accurately estimate the thermal-noise level of gravitational-wave detectors but also to help analyzing thermal noise in quantum-measurement experiments with lighter mirrors.
\\
\vspace{0.1cm}}
\vspace{1cm}
]

\section{Introduction}

Advancement in the reduction of technical noise and isolation of seismic vibration has let a high-precision measurement device like the interferometric gravitational-wave detectors~\cite{GW} be so sensitive that tiny thermal fluctuation of the measurement surface can limit the sensitivity. It is important to develop a method to estimate the thermal-noise level as accurately as possible. Our study with a finite-size mirror is an upgrade of previous works for coating thermal noise with some approximations. For mirrors currently planned to be used in the next-generation gravitational-wave detectors, the difference is a few percent between the results with our finite-size analysis and with a conventional infinite-size analysis. Besides, our analysis, for the first time, includes the effect of temperature fluctuations in the substrate and the coatings coherently summed up, with which the estimation of thermal noise will be more accurate at lower frequencies. The difference from the previous results can be larger if the mirror is thin. Thermal noise is also important in cold damping experiments~\cite{cold}, where the purpose is to reach a quantum limit with a low-mass mirror, which may tend to be thin. In this paper, we show calculation results of coating thermal noise with a broad range of aspect ratios, which agree to the previous results with an infinite-size mirror in the thick limit, and also agree to the results with a thin-plate that are calculated using the modal-expansion method. It is important to know the thermal-noise level in the middle range so that an appropriate mirror can be used in the various experiments.

There are two different ways that non-zero temperature causes fluctuation of the surface of a mirror. The first one is via volume fluctuation under fixed temperature; called Brownian thermal noise. Brownian thermal noise in the power spectrum density (m/$\sqrt{\mathrm{Hz}}$) is proportional to the square-root of temperature $\sqrt{T}$, besides the mechanical loss has some temperature dependence. The second one is via temperature fluctuation that converts into the surface fluctuation through the thermal expansion and through the change of the refraction index. Thermal noise through the expansion is called thermoelastic noise~\cite{BraginskyTE} and thermal noise through the change of the refraction index is called thermorefractive noise~{\cite{BraginskyTRc}\cite{BraginskyTR}}; the coherent sum of thermoelastic noise and thermorefractive noise is called thermo-optic noise in Ref.~\cite{Matt}. Thermo-optic noise in the power spectrum density is linearly proportional to $T$, besides some parameters like the thermal conductivity or the thermal expansion depend on the temperature.

Brownian thermal noise is related to the mechanical loss angle $\phi$. A current gravitational-wave detector employs a mirror made of silica coated by tantala-silica doublets, and the loss angle of the silica substrate is several orders lower than that of the coatings~{\cite{Penn}\cite{Harry}}. Thermo-optic noise is related to the heat flow in the $r$-direction (transverse to the beam) and in the $z$-direction (along the beam) of the cylindrical mirror. The both contributions are to be taken into account in the case without coatings~\cite{Liu}, while the latter becomes dominant with coatings according to the difference of the mechanical parameters of the materials~\cite{Fejer}. In this paper, we focus on the derivation of Brownian thermal noise in the coatings and thermo-optic noise in the $z$-direction.

Historically, Brownian thermal noise of a mirror had been analyzed using a so-called modal-expansion method~\cite{Saulson}. Gillespie and Raab demonstrated a calculation with Hutchinson's method to derive the contribution from each mechanical mode of an axisymmetric cylinder~\cite{Raab}. The contributions are added up with a weight function given by the power distribution of the Gaussian beam that probes the mirror. In 1998, Levin proposed a new way to analyze thermal noise using fluctuation-dissipation theorem~{\cite{Levin}\cite{LevinPLA}}. Thermal noise is given from the multiple of the loss angle and the elastic energy of a mirror imaginarily pushed by the Gaussian beam. Levin demonstrated a calculation for substrate thermal noise with an approximation that the mirror is an infinite half-space, which is reasonable as the beam size on the mirror is usually set sufficiently small compared with the mirror radius to avoid the diffractive loss. Bondu {\it et al} calculated substrate thermal noise of a finite cylinder using Levin's method~\cite{Bondu}. The elastic energy of a monolithic substrate was derived with the boundary conditions of a finite cylinder. Harry {\it et al} extended the elastic equation with coating layers on the substrate and calculated Brownian thermal noise of thin coatings on an infinite mirror. In this paper, we will derive Brownian thermal noise of coatings on a finite cylinder.

Thermoelastic noise, introduced in Ref.~\cite{Zener}, is related to the thermal expansion, and the fluctuation-dissipation theorem is again useful to derive the expansion. Liu and Thorne calculated thermoelastic noise of an uncoated substrate associated with the heat flow both in the $r$- and $z$- directions using Levin's method~\cite{Liu}. Thermoelastic noise of a coated material by the heat flow in the $z$-direction was calculated by Braginsky {\it et al} with an approximation that thermoelastic dissipation via non-zero relaxation time of the heat flow at the coatings be regarded instantaneous -- {\it thin-coating approximation}~\cite{BraginskyTE}. They calculated thermoelastic noise both in the infinite case and in the finite case with the thin-coating approximation. Fejer {\it et al} calculated thermoelastic noise without the thin-coating approximation; the heat equation was solved both in the substrate and in the coatings, but the mirror was an infinite half-space and the expansion was approximated to be constant in $z$~\cite{Fejer}. In this paper, we will derive thermoelastic noise by the heat flow in the $z$-direction without the thin-coating approximation and with a finite cylinder.

Thermorefractive noise is calculated, even in this paper, with the thin-coating approximation. More rigorous analysis could be done but it will require an individual treatment of each layer with as many boundary conditions as the number of layers (typically $15\sim40$), which shall be remained as a future work. Nevertheless, as will be shown in this paper, thermorefractive noise is not as sensitive to the thickness of the substrate as other two noise sources. 

The structure of this paper is as follows. In Sec.~\ref{sec:overview}, we explain the fluctuation-dissipation theorem, the elastic equation, and the heat equation. In Sec.~\ref{sec:Brownian}, we use Bondu's solution to the elastic equation with a finite cylinder and extend it with the coatings to calculate Brownian thermal noise of the coatings. In Sec.~\ref{sec:TE}, we extract the expansion term from the last result and put it into the heat equation to calculate thermoelastic noise. In Sec.~\ref{sec:TR}, we show the heat equation for thermorefractive noise, and the result will be combined with thermoelastic noise to make thermo-optic noise, which is shown in Sec.~\ref{sec:TO}. In Appendix~\ref{sec:app}, we show the results for Brownian thermal noise and thermoelastic noise calculated with the thin-plate approximation, which should agree to the results of the finite-mirror calculations in the thin limit. In the end, Appendix~\ref{sec:appB} is a list of the parameters.

\section{Overview of the method}\label{sec:overview}

\subsection{fluctuation-dissipation theorem}

The conventional modal-expansion method and Levin's method are substantially equivalent methods to derive the noise spectrum using the fluctuation-dissipation theorem. In both methods, an {\it imaginary} force is applied to the mirror. While the modal-expansion method first calculates the thermal motion of the mirror in many elastic eigenmodes and then adds them up with a weighting function for the Gaussian beam, Levin's method directly calculates the dissipation and thermal noise without the modal decomposition. In the main body of this paper, we use Levin's method. We also use the modal-expansion method in Appendix.~\ref{sec:app} to calculate thermal noise of a thin plate. Since these two methods are quite different, coincidence of the results in the thin limit validates our calculation.

The equality of the fluctuation and the dissipation is the important part of the fluctuation-dissipation theorem; the power spectrum of thermal motion is expressed by the following equation:
\begin{eqnarray}
S_x(\Omega)=\frac{4k_\mathrm{B}T}{\Omega^2}\times\mathrm{Re}[1/Z(\Omega)]\ ,
\end{eqnarray}
where $k_\mathrm{B}$ is the Boltzmann constant and $Z(\Omega)$ is the impedance of the system, which is given by
\begin{eqnarray}
Z(\Omega)=\frac{F(\Omega)}{\dot{x}(\Omega)}=\frac{F(\Omega)}{i\Omega x(\Omega)} .\label{eq:impedance}
\end{eqnarray}
The phase difference $\phi$ between the imposed force $F=F_0\cos{(\Omega t)}$ and the resulting motion $x=x_0\cos{(\Omega t-\phi)}$ is called loss angle. The averaged dissipated power is the product of $F$ and $\dot{x}$ in the same phase: $W=F_0x_0\Omega\phi/2$, so, with Eq.~(\ref{eq:impedance}), the power spectrum is rewritten as
\begin{eqnarray}
S_x(\Omega)=\frac{8k_\mathrm{B}TW}{\Omega^2 F_0^2}\ \left(=\frac{8k_\mathrm{B}T}{\Omega F_0^2}U\phi\right)\ .\label{eq:FdT}
\end{eqnarray}
Here $U$ is the maximum elastic energy that can be generated by the imaginary force. The dissipation for Brownian thermal noise is derived with the elastic equation, and the dissipation for thermoelastic noise is derived with the heat equation.

The logic above should be retraced in the case of thermorefractive noise, although Eq.~(\ref{eq:FdT}) still works as well. Thermorefractive noise is no actual motion but phase shift of the light due to the change of refraction index. Thus, it is not the imaginary force but imaginary entropy that should be added to the system to use the fluctuation-dissipation theorem. Adding the imaginary entropy, we can derive the noise spectrum of the temperature fluctuation, which can be converted into the fluctuation of the optical length. See Ref.~\cite{LevinPLA} for the detail. Note that we regard the imaginary force, or imaginary entropy, to be static. This should be good if the frequency of our interest is much lower than the resonances of a mirror.

\subsection{Elastic equation}

\begin{figure}[t]
	\begin{center}
		\includegraphics[width=8cm]{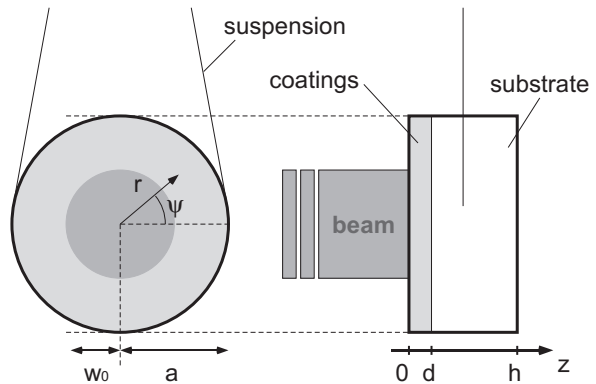}
	\caption{A cylindrical mirror.}
	\label{fig:cylinder}
	\end{center}
\end{figure}

The elastic energy is given by the product of the strain tensor $E_{ij}$ and the stress tensor $T_{ij}$, integrated over the volume of interest:
\begin{eqnarray}
U=\frac{1}{2}\int\!\sum_{i,j}E_{ij}T_{ij}dV\ \ \ \ (i,j=r,\psi,z)\ .\label{eq:U}
\end{eqnarray}
See Fig.~\ref{fig:cylinder} for the location parameters. The strain tensors of a cylinder with the axisymmetric pressure on the center are expressed by the displacement vectors $u_r$ and $u_z$ as follows:
\begin{eqnarray}
&&E_{rr}=\frac{\partial u_r}{\partial r}\ ,\ E_{\psi\psi}=\frac{u_r}{r}\ ,
E_{zz}=\frac{\partial u_z}{\partial z}\ ,\nonumber\\
&&E_{rz}=\frac{1}{2}\left(\frac{\partial u_r}{\partial z}+\frac{\partial u_z}{\partial r}\right)\ ,\label{eq:strain}
\end{eqnarray}
and the stress tensors are as follows:
\begin{eqnarray}
T_{rr}&=&(\lambda+2\mu)E_{rr}+\lambda(E_{\psi\psi}+E_{zz})\ ,\nonumber\\
T_{\psi\psi}&=&(\lambda+2\mu)E_{\psi\psi}+\lambda(E_{zz}+E_{rr})\ ,\nonumber\\
T_{zz}&=&(\lambda+2\mu)E_{zz}+\lambda(E_{rr}+E_{\psi\psi})\ ,\nonumber\\
T_{rz}&=&2\mu E_{rz}\ .\label{eq:stress}
\end{eqnarray}
Here $\lambda$ and $\mu$ are so-called Lamme coefficients:
\begin{eqnarray}
\lambda=\frac{Y\nu}{(1+\nu)(1-2\nu)}\ ,\ \ \mu=\frac{Y}{2(1+\nu)}\ ,
\end{eqnarray}
with $Y$ as Young's modulus and $\nu$ as Poisson's ratio. The other elements of the tensors are zero due to the axisymmetry. The tensors should meet Newton's second law and Hook's law. Consequently, the elastic equation in the static case is summarized into two equations:
\begin{eqnarray}
\frac{\partial^2 u_r}{\partial r^2}+\frac{\partial^2 u_r}{\partial z^2}+\frac{1}{r}\frac{\partial u_r}{\partial r}-\frac{u_r}{r^2}&=&0\ ,\label{eq:el1}\\
\frac{\partial^2 u_z}{\partial z^2}+\frac{\partial^2 u_z}{\partial r^2}+\frac{1}{r}\frac{\partial u_r}{\partial r}&=&0\ ,\label{eq:el2}
\end{eqnarray}
which are Bessel's differential equations for $r$. The solutions are given by Bondu {\it et al}~\cite{Bondu} with some typos corrected by Liu and Thorne~\cite{Liu}. Boundary conditions make a difference between the solutions with an infinite-size mirror and with the finite-size mirror. We will follow their calculation for the finite-size mirror and extend it with the coatings in Sec.~\ref{sec:Brownian}.

\subsection{Heat equation}

Either by adding the imaginary force or the imaginary entropy, the imaginary heat is generated and the finite-speed heat flow results in dissipation. The heat equation is 
\begin{eqnarray}
i\Omega{\theta}_\mathrm{j}(z)-\kappa_\mathrm{j}\nabla^2\theta_\mathrm{j}(z)=q_\mathrm{j}(z)\ ,\label{eq:heatnabla}
\end{eqnarray}
where $\theta_\mathrm{j}(z)$ is the time-varying temperature that is a Fourier transform of the temperature fluctuation $\delta T(z,t)$, $\kappa_\mathrm{j}=k_\mathrm{j}/C_\mathrm{j}$ is the thermal diffusivity ($k_\mathrm{j}$ is the thermal conductivity and $C_\mathrm{j}$ is the specific heat per volume~\cite{Note}), $q_\mathrm{j}(z)$ is the heat source, and the subscript $j$ indicates substrate ($\mathrm{s}$) or coatings ($\mathrm{c}$). In the case of a coated material, the dissipation of the heat flow in the $z$-direction is larger than the $r$-direction, thus $\nabla$ in Eq.~(\ref{eq:heatnabla}) shall be replaced by $\partial/\partial z$. The heat flow $\kappa C\partial\theta/\partial z$ gives the dissipation:
\begin{eqnarray}
W&=&\frac{\kappa C}{T}\left<\int\left(\frac{\partial\delta T}{\partial z}\right)^2dV\right>\\
&=&\frac{\kappa C}{2T}\int\left|\frac{\partial\theta}{\partial z}\right|^2dV\ .\label{eq:W}
\end{eqnarray}
Here the bracket $<>$ means the time average.

The heat source $q_\mathrm{j}$ is different between the equations for thermoelastic noise and thermorefractive noise. In the case of thermoelastic noise, the heat source is the expansion due to the imaginary force and it can be derived from the law of adiabatic temperature change~{\cite{Liu}\cite{LL}}:
\begin{eqnarray}
q_\mathrm{j}^\mathrm{TE}=-i\Omega\frac{\alpha_\mathrm{j} Y_\mathrm{j} T}{C_\mathrm{j}(1-2\nu_\mathrm{j})}\Theta_\mathrm{j}\ ,
\end{eqnarray}
where $\alpha_\mathrm{j}$ is the thermal expansion, and $\Theta_\mathrm{j}$, the expansion, is expressed by the strain tensors:
\begin{eqnarray}
\Theta=E_{rr}+E_{\psi\psi}+E_{zz}\ .
\end{eqnarray}
In the previous studies~{\cite{Matt}\cite{Fejer}}, the expansion is regarded to be constant in $z$. This is good for the coatings while the $z$-dependence in the substrate, which is taken into account in our calculation, makes a non-trivial difference at low frequencies. We will explain the detail in Sec.~\ref{sec:TE}.

In the case of thermorefractive noise, the heat source is the change of the refraction index caused by the imaginary entropy perturbation~{\cite{BraginskyTR}\cite{Matt}}:
\begin{eqnarray}
q_\mathrm{c}^\mathrm{TR}=-i\Omega\frac{\beta_\mathrm{eff}\tilde{\lambda}TF_0}{C_\mathrm{c}}p(r)\delta(z)\ ,\ \ q_\mathrm{s}^\mathrm{TR}=0\ ,\label{eq:TR}
\end{eqnarray}
where $\tilde{\lambda}$ is the wavelength of light, $p(r)$ is the Gaussian profile of the beam, and $\beta_\mathrm{eff}$ is the effective temperature dependence of the refraction index given as follows:
\begin{eqnarray}
\beta_\mathrm{eff}=\frac{n_2^2\beta_1+n_1^2\beta_2}{4(n_1^2-n_2^2)}\ ,\label{eq:beta}
\end{eqnarray}
with $n_1$ and $n_2$ as the refraction indices of two coating materials ($n_1>n_2$) and $\beta_1$ and $\beta_2$ as their temperature dependence. The delta function in Eq.~(\ref{eq:TR}) means that the heat source exists in the very beginning of the coating layers. See Ref.~\cite{BraginskyTR} for the details.

As is shown in Ref.~\cite{Matt}, the heat sources through the expansion and through the change of the refraction index have opposite signs in the heat equation. Both the thermal expansion $\alpha$ and the refraction-index change $\beta$ are positive constants, i.e. the mirror expands geometrically and optically by the increase of the temperature, but the phase shifts due to the geometrical expansion and the optical expansion are opposite. Consequently, thermo-optic noise will be smaller than thermoelastic noise or thermorefractive noise alone. We will show the calculation result with a finite-size mirror in Sec.~\ref{sec:TO}.

\subsection{Mono-layer approximation}

The calculation of thermal noise in this paper, as well as other previous works, is based on the model that a single-layer coating with the thickness of multi-layer coatings is attached on a substrate. For Brownian thermal noise, the noise levels individually calculated with the silica coatings and with the tantala coatings should be square-summed. For thermoelastic noise, as is done by Fejer {\it et al}~\cite{Fejer}, we should replace some groups of parameters by the averaged value according to the following way:
\begin{eqnarray}
(X)_\mathrm{avg}=\frac{d_\mathrm{S}}{d}X_\mathrm{S}+\frac{d_\mathrm{T}}{d}X_\mathrm{T} ,
\end{eqnarray}
where $d_\mathrm{j}$ is the coating thickness of each material, $d$ is the total thickness, the subscripts $\mathrm{S}$ and $\mathrm{T}$ indicate silica and tantala, respectively, and $X$, an operator to be averaged, would be the heat source $q_\mathrm{c}$, the thermal diffusion $\kappa_\mathrm{c}$ in Eq.~(\ref{eq:heatnabla}), or the thermal conductivity $\kappa_\mathrm{c} C_\mathrm{c}$ in Eq.~(\ref{eq:W}). For thermorefractive noise, $\beta_\mathrm{eff}$ in Eq.~(\ref{eq:beta}) is already an averaged quantity. In this paper, we use a single tantala coating in Sec.~\ref{sec:TE} and an averaged coating in Sec.~\ref{sec:TO}.

We should note that using a mono-layer coating is an approximation even with the averaging. Although the probe light is after all reflected by the coatings, some fraction of it transmits through the first few layers before reflected by a later layer. More rigorous calculation would require solving the elastic equation and the heat equation with as many boundary conditions as the number of layers, and it would probably include some coherent cancellation of the volume fluctuation and the fluctuation of the refraction index. We shall leave this as a future work.

In this paper, the optical length of each coating layer is a quarter of the wavelength of the probing light. Recently Principe {\it et al} has pointed out that thermal-noise level would decrease by tuning the layer thickness~\cite{Pinto}. We shall also leave this as a future work.

\section{Brownian thermal noise}\label{sec:Brownian}

Let us follow Bondu's calculation to derive the strain and stress tensors in a cylindrical substrate. The tensors of coatings will be derived afterwards. The boundary conditions are
\begin{eqnarray}
&&T_{rz}(r,z=0)=0\ ,\ \ T_{zz}(r,z=0)=-F_0p(r)\ ,\nonumber\\
&&T_{rr}(r=a,z)=T_{rz}(r=a,z)=0\ ,\nonumber\\
&&T_{zz}(r,z=h)=T_{rz}(r,z=h)=0\ ,\label{eq:BC}
\end{eqnarray}
where
\begin{eqnarray}
p(r)=\frac{2}{\pi w_0^2}e^{-2r^2/w_0^2}
\end{eqnarray}
is the Gaussian profile of the beam. The solution to the elastic equation [Eqs.~(\ref{eq:el1})(\ref{eq:el2})] with the boundary conditions [Eq.~(\ref{eq:BC})] is 
\begin{eqnarray}
u_r&=&\sum_m{A_m(z)J_1(k_mr)}+\Delta u_r\ ,\label{eq:Am}\\
u_z&=&\sum_m{B_m(z)J_0(k_mr)}+\Delta u_z\ ,\label{eq:Bm}
\end{eqnarray}
with
\begin{eqnarray}
k_m&=&\frac{\zeta_m}{a}\ ,\nonumber\\
\frac{\Delta u_r}{F_0}&=&\frac{\lambda+2\mu}{2\mu(3\lambda+2\mu)}(c_0r+c_1rz)+\frac{\lambda p_0 r}{2\mu(3\lambda+2\mu)}\left(1-\frac{z}{h}\right)\ ,\nonumber\\
\frac{\Delta u_z}{F_0}&=&\frac{-\lambda}{\mu(3\lambda+2\mu)}\left(c_0z+\frac{c_1z^2}{2}\right)-\frac{\lambda+2\mu}{4\mu(3\lambda+2\mu)}c_1r^2\nonumber\\
&&\ -\frac{(\lambda+\mu)p_0}{\mu(3\lambda+2\mu)}\left(z-\frac{z^2}{2h}\right)+\frac{\lambda p_0 r^2}{4\mu(3\lambda+2\mu)h}\ ,\nonumber
\end{eqnarray}
and
\begin{eqnarray}
c_0=\frac{6a^2}{h^2}\sum_m{\frac{J_0(\zeta_m)p_m}{\zeta_m^2}}\ ,\ \ c_1=\frac{-2c_0}{h}\ ,\ \ p_0=\frac{1}{\pi a^2}\ ,
\end{eqnarray}
where $\zeta_m$ is a so-called Bessel-zero function that satisfies $J_1(\zeta_m)=0$, and
\begin{eqnarray}
p_m=\frac{\exp{(-k_m^2w_0^2/8)}}{\pi a^2 J_0^2(\zeta_m)}\ ,
\end{eqnarray}
which satisfies
\begin{eqnarray}
p(r)=\sum_m{p_mJ_0(k_mr)}+p_0\ .
\end{eqnarray}
In Eqs.~(\ref{eq:Am})(\ref{eq:Bm}), $A_m$ and $B_m$ are the functions of $z$:
\begin{eqnarray}
A_m(z)&=&\gamma_m e^{-k_mz}+\delta_m e^{+k_mz}\nonumber\\
&&+\frac{k_mz}{2}\frac{\lambda+\mu}{\lambda+2\mu}\left(\alpha_m e^{-k_mz}+\beta_m e^{+k_mz}\right)\ ,\nonumber\\
B_m(z)&=&\left[\frac{\lambda+3\mu}{2(\lambda+2\mu)}\alpha_m+\gamma_m\right]e^{-k_mz}\nonumber\\
&&+\left[\frac{\lambda+3\mu}{2(\lambda+2\mu)}\beta_m-\delta_m\right]e^{+k_mz}\nonumber\\
&&+\frac{k_mz}{2}\frac{\lambda+\mu}{\lambda+2\mu}\left(\alpha_m e^{-k_mz}-\beta_m e^{+k_mz}\right)\ ,\nonumber
\end{eqnarray}
with the following constants:
\begin{eqnarray}
\alpha_m&=&\frac{p_m(\lambda+2\mu)}{k_m\mu(\lambda+\mu)}\frac{1-Q_m+2k_mhQ_m}{(1-Q_m)^2-4k_m^2h^2Q_m}\ ,\nonumber\\
\beta_m&=&\frac{p_m(\lambda+2\mu)Q_m}{k_m\mu(\lambda+\mu)}\frac{1-Q_m+2k_mh}{(1-Q_m)^2-4k_m^2h^2Q_m}\ ,\nonumber\\
\gamma_m&=&-\frac{p_m}{2k_m\mu(\lambda+\mu)}\nonumber\\
&&\times\frac{[2k_m^2h^2(\lambda+\mu)+2\mu k_mh]Q_m+\mu(1-Q_m)}{(1-Q_m)^2-4k_m^2h^2Q_m}\ ,\nonumber\\
\delta_m&=&-\frac{p_m Q_m}{2k_m\mu(\lambda+\mu)}\nonumber\\
&&\times\frac{2k_m^2h^2(\lambda+\mu)-2\mu k_mh-\mu(1-Q_m)}{(1-Q_m)^2-4k_m^2h^2Q_m}\ ,\nonumber\\
Q_m&=&\exp{(-2k_mh)}\ .\nonumber
\end{eqnarray}
The derivation is shown in Ref.~{\cite{Liu}\cite{Bondu}}. Without $\Delta u_r$ and $\Delta u_z$, the solutions (\ref{eq:Am}) and (\ref{eq:Bm}) would satisfy all the boundary conditions but $T_{rr}(r=a,z)=0$. The additional terms $\Delta u_r$ and $\Delta u_z$ make the difference approximately zero. Actually these terms become dominant in the noise spectrum when we take the limit $h\ll a$.

Plugging Eqs.~(\ref{eq:Am})(\ref{eq:Bm}) into Eqs.~(\ref{eq:strain})(\ref{eq:stress}), we get the strain and stress tensors of the substrate. If we put them into Eq.~(\ref{eq:U}) and then into Eq.~(\ref{eq:FdT}), substrate thermal noise would be calculated; the result is shown in Ref.~\cite{Liu}. 

\begin{figure}[t]
	\begin{center}
		\includegraphics[width=8cm]{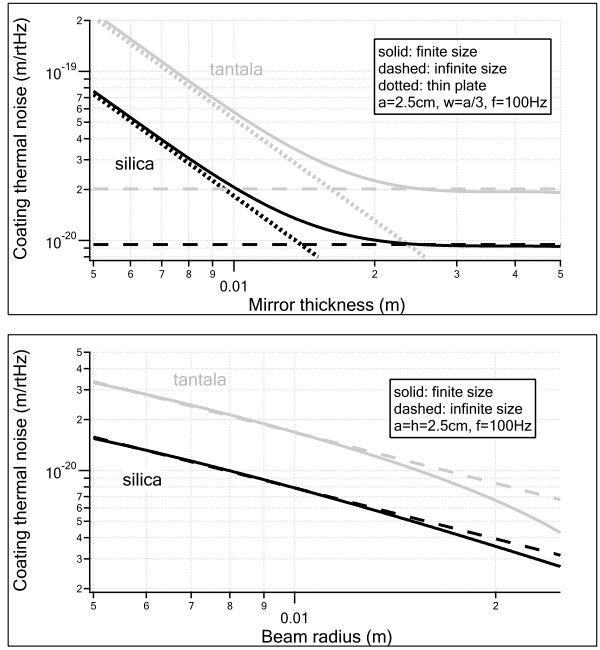}
	\caption{{\it top}: the $h$-dependence, and {\it bottom}: the $w_0$-dependence of Brownian thermal noise. They agree to the previous results in the thin limit and the thick limit.}
	\label{fig:BR}
	\end{center}
\end{figure}

With the coatings, as is introduced by Harry {\it et al}~\cite{Harry}, the boundary conditions between the substrate and the coatings are:
\begin{eqnarray}
&&E'_{rr}=E_{rr}\ ,\ \ E'_{\psi\psi}=E_{\psi\psi}\ ,\ \ E'_{rz}=E_{rz}\ ,\nonumber\\
&&T'_{zz}=T_{zz}\ ,\ \ T'_{rz}=T_{rz}\ ,
\end{eqnarray}
where the elements with a prime ($'$) are for the coatings. At the boundary and in the coatings, $E^{(')}_{rz}$ and $T^{(')}_{rz}$ are actually zero. Since the coatings are thin, we can assume that the strain and stress tensors are constant in terms of $z$. After some algebra, the strain tensor elements of the coatings are given as
\begin{eqnarray}
&&E'_{rr}=\sum_m{\frac{k_m(\gamma_m+\delta_m)}{2}\Bigl[J_0(k_mr)-J_2(k_mr)\Bigr]}\nonumber\\
&&\hspace{1cm}+\frac{(\lambda+2\mu)c_0+\lambda p_0}{2\mu(3\lambda+2\mu)}\ ,\nonumber\\
&&E'_{\psi\psi}=\sum_m{\frac{k_m(\gamma_m+\delta_m)}{2}\Bigl[J_0(k_mr)+J_2(k_mr)\Bigr]}\nonumber\\
&&\hspace{1cm}+\frac{(\lambda+2\mu)c_0+\lambda p_0}{2\mu(3\lambda+2\mu)}\ ,\nonumber\\
&&E'_{zz}=\sum_m{\biggl(\frac{-1}{\lambda'+2\mu'}k_mJ_0(k_mr)}\nonumber\\
&&\hspace{1cm}\times\Bigl[\mu(\alpha_m-\beta_m)+(\lambda'+2\mu)(\gamma_m+\delta_m)\Bigr]\biggr)\nonumber\\
&&\hspace{1cm}-\frac{\lambda'(\lambda+2\mu)c_0+(\lambda\lambda'+3\lambda\mu+2\mu^2)p_0}{\mu(3\lambda+2\mu)(\lambda'+2\mu')}\ ,\nonumber\\
&&E'_{rz}=0\ ,\nonumber\\ \label{eq:strain2}
\end{eqnarray}
then the stress tensor elements of the coatings are given as
\begin{eqnarray}
T'_{rr}&=&(\lambda'+2\mu')E'_{rr}+\lambda'(E'_{\psi\psi}+E'_{zz})\ ,\nonumber\\
T'_{\psi\psi}&=&(\lambda'+2\mu')E'_{\psi\psi}+\lambda'(E'_{zz}+E'_{rr})\ ,\nonumber\\
T'_{zz}&=&(\lambda'+2\mu')E'_{zz}+\lambda'(E'_{rr}+E'_{\psi\psi})\ ,\nonumber\\
T'_{rz}&=&0\ .\label{eq:stress2}
\end{eqnarray}
Putting these into
\begin{eqnarray}
U'=\pi\int_0^a\!\!\int_0^d\!\sum_{i,j}{E'_{ij}T'_{ij}}dzrdr\ \ (i,j=r,\psi,z)\ ,\label{eq:U2}
\end{eqnarray}
and then into Eq.~(\ref{eq:FdT}), we obtain the power spectrum of coating thermal noise. Figure~\ref{fig:BR} shows the $h$-dependence and $w_0$-dependence of the spectrum density $\sqrt{S_x(\Omega)}$. One can see that the noise level agrees to the result with an infinite-size mirror, shown by dashed curves, with $h$ sufficiently larger than $\sim a$, and increases by $h^{-2}$ as the mirror becomes thin. The dotted curves in the top panel of Fig.~\ref{fig:BR} is the result with the thin-plate calculation, which we introduce in Appendix~\ref{sec:app}. As $h$ is sufficiently smaller than $\sim a$, the results with a finite-size mirror and with a thin plate coincide. Here the frequency is 100~Hz, the mirror radius is 2.5~cm, the beam radius in the top panel is 1~cm, the thickness in the bottom panel is 2.5~cm, and the number of the coating layers is 3 for tantala and 2 for silica; these are the parameters for a quantum-measurement experiment at Hannover~\cite{10m}.

Taking the limit $a\rightarrow\infty$ and $h\rightarrow\infty$, the strain and stress tensors agree to what are shown in Ref.~\cite{Harry}, and the noise spectrum is given as
\begin{eqnarray}
&&S_x(\Omega)=\frac{4k_\mathrm{B}T}{\Omega}\frac{d}{\pi w_0^2}\nonumber\\
&&\ \ \ \ \times\frac{Y_\mathrm{c}^2(1+\nu_\mathrm{s})^2(1-2\nu_\mathrm{s})^2+Y_\mathrm{s}^2(1+\nu_\mathrm{c})^2(1-2\nu_\mathrm{c})}{Y_\mathrm{s}^2Y_\mathrm{c}(1-\nu_\mathrm{c}^2)}\phi_\mathrm{c}\ .\nonumber\\
\end{eqnarray}
For example, coating Brownian thermal noise of a mirror of a mirror in Advanced LIGO~\cite{AdLIGO},  a second-generation gravitational-wave detector ($a=17~\mathrm{cm},\ h=20~\mathrm{cm},\ w_0=6.2~\mathrm{cm}$, and with 19 doublets of silica-tantala coatings), the thermal-noise level with the finite-size analysis is $\sim2.6~\%$ times smaller than that with the infinite-size analysis.

Coating Brownian thermal noise can be numerically obtained using the finite-element method, although it takes longer time. Yamamoto {\it et al} calculated the $w_0$-dependence of coating Brownian thermal noise for a gravitational-wave detector~\cite{Yamamoto2002} and the result was identical to what we see in the bottom panel of Fig.~\ref{fig:BR}.

\section{Thermoelastic noise}\label{sec:TE}

The heat source of thermoelastic noise is the expansion. The expansion of the substrate is calculated to be
\begin{eqnarray}
&&\hspace{-0.5cm}\frac{\Theta_\mathrm{s}}{F_0}=\sum_m{(k_mA_m(z)+B'_m(z))J_0(k_mr)}\nonumber\\
&&\hspace{0.0cm}+\frac{2}{3\lambda+2\mu}(c_0+c_1z)-\frac{p_0}{3\lambda+2\mu}\left(1-\frac{z}{h}\right)\ .
\end{eqnarray}
with $B'_m(z)=dB_m/dz$. The expansion of the coatings is calculated from Eq.~(\ref{eq:strain2}) as
\begin{eqnarray}
&&\frac{\Theta_\mathrm{c}}{F_0}=\sum_m{\biggl(\frac{-1}{\lambda'+2\mu'}k_mJ_0(k_mr)}\nonumber\\
&&\hspace{0.5cm}\times\Bigl[\mu(\alpha_m-\beta_m)+2(\mu-\mu')(\gamma_m+\delta_m)\Bigr]\biggr)\nonumber\\
&&\hspace{0.5cm}+\frac{(\lambda+2\mu)2\mu'c_0+(2\lambda\mu'-3\lambda\mu-2\mu^2)p_0}{(\lambda'+2\mu')\mu(3\lambda+2\mu)}\ .\nonumber\\ \label{eq:expansion_c}
\end{eqnarray}
These expansions appear on the right side of the heat equation:
\begin{eqnarray}
i\Omega\theta_\mathrm{j}-\kappa_\mathrm{j}\frac{\partial^2}{\partial z^2}\theta_\mathrm{j}=-i\Omega\frac{\alpha_\mathrm{j} Y_\mathrm{j} T}{C_\mathrm{j}(1-2\nu_\mathrm{j})}\Theta_\mathrm{j}\ .\label{eq:heateqTE}
\end{eqnarray}
The homogeneous solution of the heat equation, which is the solution of the left-hand-side of Eq.~(\ref{eq:heateqTE}) being zero, is 
\begin{eqnarray}
\theta_\mathrm{j}^H=A_\mathrm{j}\sinh{(\gamma_\mathrm{j}z)}+B_\mathrm{j}\cosh{(\gamma_\mathrm{j}z)}\ ,\label{eq:homogeneous}
\end{eqnarray}
where $A_\mathrm{j}$ and $B_\mathrm{j}$ are coefficients that will be derived with the boundary conditions, and $\gamma_\mathrm{j}$ is the complex propagation constant given as
\begin{eqnarray}
\gamma_\mathrm{j}=(1+i)\sqrt{\frac{\Omega}{2\kappa_\mathrm{j}}}\ .
\end{eqnarray}
The particular solution of the heat equation is given to cancel the right-hand-side of Eq.~(\ref{eq:heateqTE}). Reference~\cite{Fejer} uses an approximation that both $\Theta_\mathrm{s}$ and $\Theta_\mathrm{c}$ are constant in $z$, so that the particular solution is simply
\begin{eqnarray}
\theta_\mathrm{j}^P\sim-\frac{\alpha_\mathrm{j} Y_\mathrm{j} T}{C_\mathrm{j}(1-2\nu_\mathrm{j})}\Theta_\mathrm{j}\ \ (z\simeq0)\ ,\nonumber
\end{eqnarray}
and the noise spectrum for the infinite-size mirror is described in an elegant form. In this paper, however, we shall calculate the noise spectrum without this approximation. The particular solution for the substrate without the approximation is
\begin{eqnarray}
&&\theta_\mathrm{s}^P=-\frac{\alpha_\mathrm{s} Y_\mathrm{s} TF_0}{C_\mathrm{s}(1-2\nu_\mathrm{s})}\nonumber\\
&&\hspace{0.5cm}\times\biggl[\sum_m{\frac{i\Omega}{i\Omega-\kappa_\mathrm{s} k_m^2}(k_mA_m(z)+B'_m(z))J_0(k_mr)}\nonumber\\
&&\hspace{1cm}+\frac{2}{3\lambda+2\mu}(c_0+c_1z)-\frac{p_0}{3\lambda+2\mu}\left(1-\frac{z}{h}\right)\biggr]\ .\nonumber\\ \label{eq:particularS}
\end{eqnarray}
The particular solution for the coatings is simply
\begin{eqnarray}
\theta_\mathrm{c}^P=-\frac{\alpha_\mathrm{c} Y_\mathrm{c} TF_0}{C_\mathrm{c}(1-2\nu_\mathrm{c})}\Theta_\mathrm{c}\ ,\label{eq:particularC}
\end{eqnarray}
with $\Theta_\mathrm{c}$ in Eq.~(\ref{eq:expansion_c}). 

The complete solution is the sum of the homogeneous and particular solutions:
\begin{eqnarray}
\theta_\mathrm{j}=\theta_\mathrm{j}^H+\theta_\mathrm{j}^P\ .\label{eq:complete}
\end{eqnarray}
There are four boundary conditions to be met, which are (i) the heat flow at $z=0$ is zero, (ii) the heat flow at $z=h$ is zero, (iii) the heat flows from the coatings to the substrate and from the substrate to the coatings are equal, and (iv) the temperature at the border is also equal:
\begin{eqnarray}
&&(\mathrm{i})\ \left.\kappa_\mathrm{c}C_\mathrm{c}\frac{\partial\theta_\mathrm{c}}{\partial z}\right|_{z=0}=0\ ,\label{eq:BC1}\\
&&(\mathrm{ii})\ \left.\kappa_\mathrm{s}C_\mathrm{s}\frac{\partial\theta_\mathrm{s}}{\partial z}\right|_{z=h}=0\ ,\label{eq:BC2}\\
&&(\mathrm{iii})\ \left.\kappa_\mathrm{c}C_\mathrm{c}\frac{\partial\theta_\mathrm{c}}{\partial z}\right|_{z=d}=\left.\kappa_\mathrm{s}C_\mathrm{s}\frac{\partial\theta_\mathrm{s}}{\partial z}\right|_{z=d}\ ,\label{eq:BC3}\\
\vspace{0.5cm}\nonumber\\
&&(\mathrm{iv})\ \ \theta_\mathrm{c}=\theta_\mathrm{s}(d)\label{eq:BC4}\ .
\end{eqnarray}
The coefficients in the homogeneous solution are then given as
\begin{eqnarray}
&&\hspace{-0.5cm}A_\mathrm{c}=0\ ,\nonumber\\
&&\hspace{-0.5cm}B_\mathrm{c}\simeq\frac{\Sigma_1/\gamma_\mathrm{s}+(\Sigma_2-\theta_\mathrm{c}^P)}{\cosh{(\gamma_\mathrm{c} d)}+R\sinh{(\gamma_\mathrm{c} d)}}\ ,\nonumber\\
&&\hspace{-0.5cm}A_\mathrm{s}\simeq\frac{-(\Sigma_1/\gamma_\mathrm{s})\cosh{(\gamma_\mathrm{c} d)}+(\Sigma_2-\theta_\mathrm{c}^P)R\sinh{(\gamma_\mathrm{c} d)}}{\cosh{(\gamma_\mathrm{c} d)}+R\sinh{(\gamma_\mathrm{c} d)}}e^{\gamma_\mathrm{s} d}\ ,\nonumber\\
&&\hspace{-0.5cm}B_\mathrm{s}\simeq-A_\mathrm{s}\ ,\label{eq:ABs}
\end{eqnarray}
where
\begin{eqnarray}
R=\frac{\kappa_\mathrm{c}C_\mathrm{c}\gamma_\mathrm{c}}{\kappa_\mathrm{s}C_\mathrm{s}\gamma_\mathrm{s}}=\frac{\sqrt{\kappa_\mathrm{c}}C_\mathrm{c}}{\sqrt{\kappa_\mathrm{s}}C_\mathrm{s}}\ ,
\end{eqnarray}
and also $\Sigma_1=\partial\theta_\mathrm{s}^P/\partial z|_{z=d}\simeq\partial\theta_\mathrm{s}^P/\partial z|_{z=0}$ and $\Sigma_2=\theta_\mathrm{s}^P(d)\simeq\theta_\mathrm{s}^P(0)$; namely,
\begin{eqnarray}
&&\hspace{-0.5cm}\Sigma_1\simeq\frac{\alpha_\mathrm{s} Y_\mathrm{s} T}{C_\mathrm{s}(1-2\nu_\mathrm{s})}\biggl[\sum_m{\frac{-i\Omega k_m^2}{i\Omega-\kappa_\mathrm{s} k_m^2}\frac{\mu(\alpha_m+\beta_m)}{\lambda+2\mu}J_0(k_mr)}\nonumber\\
&&\hspace{0.5cm}-\frac{1}{3\lambda+2\mu}\left(2c_1+\frac{p_0}{h}\right)\biggr]\ ,\nonumber\\
&&\hspace{-0.5cm}\Sigma_2\simeq\frac{\alpha_\mathrm{s} Y_\mathrm{s} T}{C_\mathrm{s}(1-2\nu_\mathrm{s})}\biggl[\sum_m{\frac{i\Omega k_m}{i\Omega-\kappa_\mathrm{s} k_m^2}\frac{\mu(\alpha_m-\beta_m)}{\lambda+2\mu}J_0(k_mr)}\nonumber\\
&&\hspace{0.5cm}-\frac{1}{3\lambda+2\mu}\left(2c_0-p_0\right)\biggr]\ .\label{eq:Sigmas}
\end{eqnarray}
Here we use two approximations. One is to ignore the terms with $e^{-\gamma_\mathrm{s} h}$ in the presence of other terms in Eq.~(\ref{eq:ABs}); this is fine as far as the target frequency is higher than the inverse of the relaxation time of the temperature gradient. The other is to ignore the difference between $e^{\pm k_m d}$ and unity as well as to ignore the terms with $d/h$ in Eq.~(\ref{eq:Sigmas}); this is fine if the beam radius is not as small as the order of $d$. Just in case, the followings are the terms that could be added to each term in Eq.~(\ref{eq:ABs}):
\begin{eqnarray}
&&\hspace{-0.5cm}\tilde{B}_\mathrm{c}=-\frac{e^{-\gamma_\mathrm{s} (h-d)}\times[\partial\theta_\mathrm{s}^P/\partial z|_{z=h}]}{\gamma_\mathrm{s}[\cosh{(\gamma_\mathrm{c} d)}+R\sinh{(\gamma_\mathrm{c} d)}]}\ ,\nonumber\\
&&\hspace{-0.5cm}\tilde{A}_\mathrm{s}=\tilde{B}_\mathrm{c}[\sinh{(\gamma_\mathrm{s} d)}\cosh{(\gamma_\mathrm{c} d)}-R\cosh{(\gamma_\mathrm{s} d)}\sinh{(\gamma_\mathrm{c} d)}]\ ,\nonumber\\
&&\hspace{-0.5cm}\tilde{B}_\mathrm{s}=\tilde{B}_\mathrm{c}[\cosh{(\gamma_\mathrm{s} d)}\cosh{(\gamma_\mathrm{c} d)}-R\sinh{(\gamma_\mathrm{s} d)}\sinh{(\gamma_\mathrm{c} d)}]\ .\nonumber\\ \label{eq:approximation1}
\end{eqnarray}
Plugging the coefficients in Eq.~(\ref{eq:ABs}) into the homogeneous solution [Eq.~(\ref{eq:homogeneous})], adding the particular solution [Eqs.~(\ref{eq:particularS})(\ref{eq:particularC})], and then putting $\theta_\mathrm{s}$ and $\theta_\mathrm{c}$ [Eq.~(\ref{eq:complete})] into the following equation:
\begin{eqnarray}
&&W=2\pi\frac{\kappa_\mathrm{c} C_\mathrm{c}}{2T}\int_0^a\!\!\int_0^d\!\left|\frac{\partial\theta_\mathrm{c}}{\partial z}\right|^2\!dzrdr\nonumber\\
&&\hspace{2.0cm}+2\pi\frac{\kappa_\mathrm{s} C_\mathrm{s}}{2T}\int_0^a\!\!\int_d^h\!\left|\frac{\partial\theta_\mathrm{s}}{\partial z}\right|^2\!dzrdr\ ,\label{eq:WTE}
\end{eqnarray}
we obtain the thermoelastic dissipating power, which with Eq.~(\ref{eq:FdT}) gives the power spectrum of thermoelastic noise.

Taking the limit $a\rightarrow\infty$ and $h\rightarrow\infty$, and ignoring the $z$-dependence of the expansion in the substrate, the noise spectrum agrees to the result of Ref.~\cite{Fejer}:
\begin{eqnarray}
S_x(\Omega)=\frac{16k_\mathrm{B} T^2d(1+\nu_\mathrm{s})^2\alpha_\mathrm{s}^2 C_\mathrm{c}}{\pi C_\mathrm{s}^2w_0^2\Omega}\tilde{\Delta}^2(g_1+g_2)\ ,\label{eq:Fejer}
\end{eqnarray}
with
\begin{eqnarray}
&&\hspace{-0.5cm}\tilde{\Delta}=\frac{\alpha_\mathrm{c} C_\mathrm{s}}{2\alpha_\mathrm{s} C_\mathrm{c}}\frac{1}{1-\nu_\mathrm{c}}\left(\frac{1+\nu_\mathrm{c}}{1+\nu_\mathrm{s}}+(1-2\nu_\mathrm{s})\frac{Y_\mathrm{c}}{Y_\mathrm{s}}\right)-1\ ,\nonumber\\ \label{eq:Fejer}
&&\hspace{-0.5cm}g_1=(\sinh{\xi}-\sin{\xi})/(\xi\Gamma_D)\ ,\nonumber\\
&&\hspace{-0.5cm}g_2=R(\cosh{\xi}-\cos{\xi})/(\xi\Gamma_D)\ ,\nonumber\\
&&\hspace{-0.5cm}\Gamma_D=(1+R^2)\cosh{\xi}+(1-R^2)\cos{\xi}+2R\sinh{\xi}\ ,\nonumber\\ \label{eq:GammaD}
&&\hspace{-0.5cm}\xi=\sqrt{2\Omega d^2/\kappa_\mathrm{c}}\ .
\end{eqnarray}
The noise spectrum with $g_1$ is the contribution of the dissipation in the coatings [the first term in Eq.~(\ref{eq:WTE})] and that with $g_2$ is the contribution of the dissipation in the substrate [the second term in Eq.~(\ref{eq:WTE})].

\begin{figure}[t]
	\begin{center}
		\includegraphics[width=8cm]{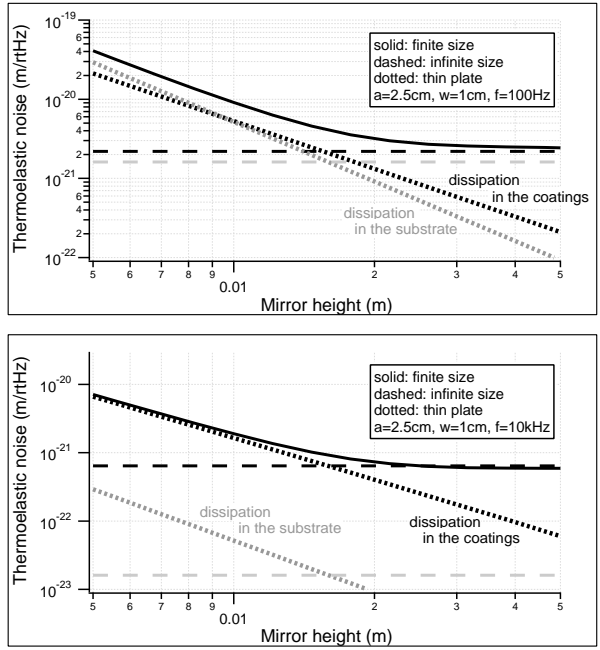}
	\caption{Thermoelastic noise at 100~Hz (top panel) and at 10~kHz (bottom panel). The solid curve is thermoelastic noise of a finite-size mirror caused by the heat both in the coatings and the substrate. The black dotted curve is thermoelastic noise caused by the heat in the coatings and the gray dotted curve is the one by the heat in the substrate; both are with the thin-plate calculation. The dashed curves are thermoelastic noise of an infinite-size mirror. Substrate thermoelastic noise is calculated only from the dissipation of the heat flow in the $z$-direction.}
	\label{fig:TE}
	\end{center}
\end{figure}

A difference between our result and the result in Ref.~\cite{Fejer} according to the $z$-dependence of the expansion in the substrate could be compensated by adding substrate thermoelastic noise of an infinite-size mirror derived with the thin-coating approximation in Ref.~\cite{BraginskyTE}:
\begin{eqnarray}
S_x(\Omega)=\frac{16k_\mathrm{B}T^2(1+\nu_\mathrm{s})^2\alpha_\mathrm{s}^2\kappa_\mathrm{s}}{\sqrt{\pi}C_\mathrm{s}w_0^3\Omega^2}\ .\label{eq:subTE}
\end{eqnarray}
Here, we use the term substrate thermoelastic noise as what is caused by the heat  source in the substrate due to the $z$-dependence of the expansion, while coating thermoelastic noise is by the heat source at the interface of the substrate and the coatings due to the difference of the materials. Note that both are contained in the $\theta_\mathrm{s}$ term and are dissipated in the substrate. Therefore, those two kinds of thermoelastic noise in the substrate should be coherently summed up and our calculation offers a proper treatment by taking into account the $z$-dependence of the expansion in the substrate. Equation~(\ref{eq:subTE}) is derived from the dissipation of the heat flow both in the $r$- and $z$-directions. Since our calculation contains only that in the $z$-direction, which is the more significant one for coating thermoelastic noise, we shall resolve Eq.~(\ref{eq:subTE}) into two parts. Following the derivation of substrate thermoelastic noise shown in Ref.~\cite{Liu}, we can easily find out that the dissipation of the heat flow in the $r$-direction and that in the $z$-direction are equal, thus
\begin{eqnarray}
S_x^{(r)}(\Omega)=S_x^{(z)}(\Omega)=\frac{8k_\mathrm{B}T^2(1+\nu_\mathrm{s})^2\alpha_\mathrm{s}^2\kappa_\mathrm{s}}{\sqrt{\pi}C_\mathrm{s}w_0^3\Omega^2}\ .\label{eq:subTErz}
\end{eqnarray}

Figure~\ref{fig:TE} shows the $h$-dependence of thermoelastic noise at two different frequencies. In the top panel, at 100~Hz, as $h$ becomes small, the result with a finite-size mirror coincides to the level of substrate thermoelastic noise of a thin plate. We can see a non-trivial difference in a broad middle range between the result with our finite-size analysis and previous results with the approximations. In the bottom panel, at 10~kHz, as $h$ becomes small, the result with a finite-size mirror coincides to the level of coating thermoelastic noise of a thin plate. The result with a finite-size mirror coincides to the result with an infinite-size mirror as $h$ becomes large. Substrate thermoelastic noise contributes more when the measurement frequency is low, the mirror is thin, and/or the coating is less. Note that the noise level with the finite-size analysis is larger than the square-sum of coating thermoelastic noise and substrate thermoelastic noise due to their correlation.

\section{Thermorefractive noise}\label{sec:TR}

The heat source of thermorefractive noise is the change of refraction index. The heat equation for thermorefractive noise is
\begin{eqnarray}
&&i\Omega\theta_\mathrm{c}-\kappa_\mathrm{c}\frac{\partial^2}{\partial z^2}\theta_\mathrm{c}=-i\Omega\frac{\beta_\mathrm{eff}\tilde{\lambda}TF_0}{C_\mathrm{c}}\cdot\frac{e^{-z/\ell}}{\ell}p(r)\ ,\nonumber\\
&&i\Omega\theta_\mathrm{s}-\kappa_\mathrm{s}\frac{\partial^2}{\partial z^2}\theta_\mathrm{s}=0\ ,\label{eq:heateqTR}
\end{eqnarray}
where we have replaced the delta function to $e^{-z/\ell}$ with an arbitrary small length $\ell$, which does not appear in the final result as far as it is sufficiently small. The boundary conditions with a finite-size mirror are same as Eqs.~(\ref{eq:BC1})-(\ref{eq:BC4}). The particular solution is
\begin{eqnarray}
\theta_\mathrm{c}^P&=&\frac{-i\Omega}{i\Omega-\kappa_\mathrm{c}/\ell^2}\frac{\beta_\mathrm{eff}\tilde{\lambda}TF_0}{C_\mathrm{c}}\frac{e^{-z/\ell}}{\ell}p(r)\nonumber\\
&\simeq&\frac{i\Omega \beta_\mathrm{eff}\tilde{\lambda}TF_0}{\kappa_\mathrm{c}C_\mathrm{c}/\ell}e^{-z/\ell}p(r)\ ,\nonumber\\
\theta_\mathrm{s}^P&=&0\ ,
\end{eqnarray}
and then the coefficients of the homogeneous solution, which is in the same form as in Eq.~(\ref{eq:homogeneous}), are derived:
\begin{eqnarray}
&&A_\mathrm{c}=i\Omega\frac{\beta_\mathrm{eff}\tilde{\lambda}TF_0}{\kappa_\mathrm{c} C_\mathrm{c}\gamma_\mathrm{c}}p(r)\ ,\nonumber\\
&&B_\mathrm{c}=-i\Omega\frac{\beta_\mathrm{eff}\tilde{\lambda}TF_0}{\kappa_\mathrm{c} C_\mathrm{c}\gamma_\mathrm{c}}p(r)\frac{\eta_1 R\cosh{(\gamma_\mathrm{c} d)}+\eta_2 \sinh{(\gamma_\mathrm{c} d)}}{\eta_1 R\sinh{(\gamma_\mathrm{c} d)}+\eta_2 \cosh{(\gamma_\mathrm{c} d)}}\ ,\nonumber\\
&&A_\mathrm{s}=i\Omega\frac{\beta_\mathrm{eff}\tilde{\lambda}TF_0}{\kappa_\mathrm{c} C_\mathrm{c}\gamma_\mathrm{c}}p(r)\frac{R\eta_0}{\eta_2 \cosh{(\gamma_\mathrm{c} d)}+\eta_1 R\sinh{(\gamma_\mathrm{c} d)}}\ ,\nonumber\\
&&B_\mathrm{s}=-\frac{1}{\eta_0} A_\mathrm{s} ,\label{eq:ABsTR}
\end{eqnarray}
where
\begin{eqnarray}
\eta_0&=&\frac{1-\exp{(-2\gamma_\mathrm{s} h)}}{1+\exp{(-2\gamma_\mathrm{s} h)}}\ \ \ \ (\simeq1)\ ,\nonumber\\
\eta_1&=&\cosh{(\gamma_\mathrm{s} d)}-\eta_0\sinh{(\gamma_\mathrm{s} d)}\ ,\nonumber\\
\eta_2&=&\eta_0\cosh{(\gamma_\mathrm{s} d)}-\sinh{(\gamma_\mathrm{s} d)}\ .\label{eq:approximation2}
\end{eqnarray}
As $\eta_0\simeq1$, thus $\eta_1\simeq\eta_2$, Eq.~(\ref{eq:ABsTR}) can be more simplified. In fact, after the simplification, none of the coefficients contains $h$. The thermorefractive-noise level with a finite-size mirror is then almost same as that with an infinite-size mirror:
\begin{eqnarray}
S_x(\Omega)=\frac{2\sqrt{2}k_\mathrm{B}T^2}{\sqrt{\Omega}}\frac{\Gamma_2}{\Gamma_D}\frac{1}{\sqrt{\kappa_\mathrm{c}}Cc}\frac{1}{\pi w_0^2}\beta_\mathrm{eff}^2\tilde{\lambda}^2\ ,
\end{eqnarray}
where 
\begin{eqnarray}
\Gamma_2=(1+R^2)\sinh{\xi}+(1-R^2)\sin{\xi}+2R\cosh{\xi}\ ,\label{eq:Gamma2}
\end{eqnarray}
and $\Gamma_D$ has been given. We introduce $\Gamma_2$ earlier than $\Gamma_0$ and $\Gamma_1$, which will be shown in Sec~\ref{sec:TO}, in order to keep the same notation as Ref.~\cite{Matt}.

\section{Thermo-optic noise}\label{sec:TO}

It has been pointed out by Evans {\it et al} that thermoelastic noise and thermorefractive noise should be coherently added with a proper treatment due to their common origin; it is now called thermo-optic noise~\cite{Matt}. In Sec.~\ref{sec:TE} and Sec.~\ref{sec:TR}, we have studied the behavior of these two kinds of noise and derived the individual noise levels, but what should be used for the noise estimation is the result in this section. The heat equation of thermo-optic noise has both thermoelastic heat source and thermorefractive heat source on the right-hand side:
\begin{eqnarray}
&&i\Omega\theta_\mathrm{c}-\kappa_\mathrm{c}\frac{\partial^2}{\partial z^2}\theta_\mathrm{c}\nonumber\\
&&\hspace{0.5cm}=-i\Omega\frac{\beta_\mathrm{eff}\tilde{\lambda}TF_0}{C_\mathrm{c}}\cdot\frac{e^{-z/\ell}}{\ell}p(r)-i\Omega\frac{\alpha_\mathrm{c} Y_\mathrm{c} T}{C_\mathrm{c}(1-2\nu_\mathrm{c})}\Theta_\mathrm{c}\ ,\nonumber\\
&&i\Omega\theta_\mathrm{s}-\kappa_\mathrm{s}\frac{\partial^2}{\partial z^2}\theta_\mathrm{s}=-i\Omega\frac{\alpha_\mathrm{s} Y_\mathrm{s} T}{C_\mathrm{s}(1-2\nu_\mathrm{s})}\Theta_\mathrm{s}\ .\label{eq:heateqTO}
\end{eqnarray}
The homogeneous solution is in the same form as Eq.~(\ref{eq:homogeneous}). The particular solution for the coatings is
\begin{eqnarray}
\theta_\mathrm{c}^P=\frac{i\Omega\beta_\mathrm{eff}\tilde{\lambda}TF_0}{\kappa_\mathrm{c}C_\mathrm{c}/\ell}e^{-z/\ell}p(r)-\frac{\alpha_\mathrm{c}Y_\mathrm{c}T}{C_\mathrm{c}(1-2\nu_\mathrm{c})}\Theta_\mathrm{c}\ ,
\end{eqnarray}
and the particular solution for the substrate is same as what we have derived for thermoelastic noise [Eq.~(\ref{eq:particularS})]. The boundary conditions with a finite-size mirror are same as Eqs.~(\ref{eq:BC1})-(\ref{eq:BC4}). Let us use the approximations (\ref{eq:approximation1}) and (\ref{eq:approximation2}), which have been proven to be safe in the individual calculations for thermoelastic noise and thermorefractive noise. The coefficients of the homogeneous solution are then given as
\begin{eqnarray}
&&\hspace{-0.5cm}A_\mathrm{c}=i\Omega\frac{\beta_\mathrm{eff}\tilde{\lambda}TF_0}{\kappa_\mathrm{c} C_\mathrm{c}\gamma_\mathrm{c}}p(r)\ ,\nonumber\\
&&\hspace{-0.5cm}B_\mathrm{c}\simeq\frac{\Sigma_1/\gamma_\mathrm{s}+(\Sigma_2+\Pi_1)}{\cosh{(\gamma_\mathrm{c} d)}+R\sinh{(\gamma_\mathrm{c} d)}}\ ,\nonumber\\
&&\hspace{-0.5cm}A_\mathrm{s}\simeq\frac{-(\Sigma_1/\gamma_\mathrm{s})\cosh{(\gamma_\mathrm{c} d)}+(\Sigma_2+\Pi_2)R\sinh{(\gamma_\mathrm{c} d)}}{\cosh{(\gamma_\mathrm{c} d)}+R\sinh{(\gamma_\mathrm{c} d)}}e^{\gamma_\mathrm{s} d}\ ,\nonumber\\
&&\hspace{-0.5cm}B_\mathrm{s}\simeq-A_\mathrm{s}\ ,\label{eq:ABsTO}
\end{eqnarray}
where
\begin{eqnarray}
&&\hspace{-0.3cm}\Pi_1=\frac{\alpha_\mathrm{c}Y_\mathrm{c}T}{C_\mathrm{c}(1-2\nu_\mathrm{c})}\Theta_\mathrm{c}\nonumber\\
&&\hspace{0.5cm}-\frac{i\Omega\beta_\mathrm{eff}\tilde{\lambda}TF_0}{\kappa_\mathrm{c}C_\mathrm{c}\gamma_\mathrm{c}}p(r)\left[\sinh{(\gamma_\mathrm{c}d)}+R\cosh{(\gamma_\mathrm{c}d)}\right]\ ,\nonumber\\
&&\hspace{-0.3cm}\Pi_2=\frac{\alpha_\mathrm{c}Y_\mathrm{c}T}{C_\mathrm{c}(1-2\nu_\mathrm{c})}\Theta_\mathrm{c}+\frac{i\Omega\beta_\mathrm{eff}\tilde{\lambda}TF_0}{\kappa_\mathrm{c}C_\mathrm{c}\gamma_\mathrm{c}\sinh{(\gamma_\mathrm{c} d)}}p(r)\ .\label{eq:Pi}
\end{eqnarray}
One can see that the coefficients in Eq.~(\ref{eq:ABsTO}) coincide to those in Eq.~(\ref{eq:ABs}) if a thermorefractive constant $\beta_\mathrm{eff}$ is supposedly erased and they coincide to those in Eq.~(\ref{eq:ABsTR}) if thermoelastic constants $\alpha_\mathrm{c}$ and $\alpha_\mathrm{s}$ are erased. 

As well as we have done in the previous sections, plugging $\theta_\mathrm{s}$ and $\theta_\mathrm{c}$ with the coefficients in Eq.~(\ref{eq:ABsTO}) into the following equation:
\begin{eqnarray}
&&W=2\pi\frac{\kappa_\mathrm{c} C_\mathrm{c}}{2T}\int_0^a\!\!\int_0^d\!\left|\frac{\partial\theta_\mathrm{c}}{\partial z}\right|^2\!dzrdr\nonumber\\
&&\hspace{2.0cm}+2\pi\frac{\kappa_\mathrm{s} C_\mathrm{s}}{2T}\int_0^a\!\!\int_d^h\!\left|\frac{\partial\theta_\mathrm{s}}{\partial z}\right|^2\!dzrdr\ ,\nonumber
\end{eqnarray}
we obtain the thermo-optic dissipating power, which with Eq.~(\ref{eq:FdT}) gives the power spectrum of thermoelastic noise. Taking the limit $a\rightarrow\infty$ and $h\rightarrow\infty$, and ignoring the $z$-dependence of the expansion in the substrate, the noise spectrum agrees to the result of Ref.~\cite{Matt}:
\begin{eqnarray}
&&\hspace{-0.8cm}S_x(\Omega)=\frac{2k_\mathrm{B}T^2}{\Omega}\frac{1}{\pi w_0^2}\frac{1}{C_\mathrm{c}}\frac{1}{\xi d}\frac{1}{\Gamma_D}\nonumber\\
&&\hspace{0cm}\times\biggl[
\Gamma_0\left(\Delta\alpha d\right)^2
-\Gamma_1\Delta\alpha d\cdot
\beta_\mathrm{eff}\tilde{\lambda}\xi
+\Gamma_2(\beta_\mathrm{eff}\tilde{\lambda}\xi)^2
\biggr]\ ,\nonumber\\ \label{eq:SxMatt}
\end{eqnarray}
where
\begin{eqnarray}
\Delta\alpha=2\alpha_\mathrm{s}(1+\nu_\mathrm{s})\frac{C_\mathrm{c}}{C_\mathrm{s}}\tilde{\Delta}\ ,
\end{eqnarray}
and
\begin{eqnarray}
\Gamma_0&=&2(\sinh{\xi}-\sin{\xi})+2R(\cosh{\xi}-\cos{\xi})\ ,\nonumber\\
\Gamma_1&=&8\sin{\frac{\xi}{2}}\left(R\cosh{\frac{\xi}{2}}+\sinh{\frac{\xi}{2}}\right)\ .
\end{eqnarray}
See Eq.~(\ref{eq:GammaD}) for $\tilde{\Delta}$, and Eqs.~(\ref{eq:GammaD})(\ref{eq:Gamma2}) for $\Gamma_D$ and $\Gamma_2$, respectively.

With the averaging, some of the constants in Eq.~(\ref{eq:SxMatt}) should be replaced as follows:
\begin{eqnarray}
\Delta\alpha\rightarrow \Delta\bar{\alpha}=\bar{\alpha}_\mathrm{c}-2\alpha_\mathrm{s}(1+\nu_\mathrm{s})\frac{C_\mathrm{c}}{C_\mathrm{s}}\ ,
\end{eqnarray}
with
\begin{eqnarray}
\bar{\alpha}_\mathrm{c}=\left[\alpha_\mathrm{c}\frac{1+\nu_\mathrm{s}}{1-\nu_\mathrm{c}}\left(\frac{1+\nu_\mathrm{c}}{1+\nu_\mathrm{s}}+(1-2\nu_\mathrm{s})\frac{Y_\mathrm{c}}{Y_\mathrm{s}}\right)\right]_\mathrm{avg}\ ,
\end{eqnarray}
and
\begin{eqnarray}
&&C_\mathrm{c}\rightarrow \bar{C}_\mathrm{c}=C_S\frac{d_S}{d}+C_T\frac{d_T}{d}\ ,\nonumber\\
&&\kappa_\mathrm{c}\rightarrow \bar{\kappa}_\mathrm{c}=\frac{1}{\bar{C}_\mathrm{c}}\left[\frac{1}{\kappa_SC_S}\frac{d_S}{d}+\frac{1}{\kappa_TC_T}\frac{d_T}{d}\right]^{-1}\ ,
\end{eqnarray}
then $\xi$, $\gamma_\mathrm{c}$, and $R$ should be replaced to the averaged constants with these new $C_\mathrm{c}$ and $\kappa_\mathrm{c}$.

The averaging can be done for the calculation with a finite-size mirror as well. With a proper treatment, $\Pi_1$ and $\Pi_2$ are replaced to
\begin{eqnarray}
&&\hspace{-0.3cm}\bar{\Pi}_1=\frac{1}{C_\mathrm{c}}\left[\frac{\alpha_\mathrm{c}Y_\mathrm{c}T}{1-2\nu_\mathrm{c}}\Theta_\mathrm{c}\right]_\mathrm{avg}\nonumber\\
&&\hspace{0.5cm}-\frac{i\Omega\beta_\mathrm{eff}\tilde{\lambda}TF_0}{\kappa_\mathrm{c}C_\mathrm{c}\gamma_\mathrm{c}}p(r)\left[\sinh{(\gamma_\mathrm{c}d)}+R\cosh{(\gamma_\mathrm{c}d)}\right]\ ,\nonumber\\
&&\hspace{-0.3cm}\bar{\Pi}_2=\frac{1}{C_\mathrm{c}}\left[\frac{\alpha_\mathrm{c}Y_\mathrm{c}T}{1-2\nu_\mathrm{c}}\Theta_\mathrm{c}\right]_\mathrm{avg}+\frac{i\Omega\beta_\mathrm{eff}\tilde{\lambda}TF_0}{\kappa_\mathrm{c}C_\mathrm{c}\gamma_\mathrm{c}\sinh{(\gamma_\mathrm{c} d)}}p(r)\ .\nonumber\\ \label{eq:PiAvg}
\end{eqnarray}
Replacing $C_\mathrm{c}$, $\kappa_\mathrm{c}$, $\xi$, $\gamma_\mathrm{c}$, and $R$ to the averaged constants shown above as well, we obtain the noise spectrum with the multi-layer coatings of adequate accuracy.

\begin{figure}[t]
	\begin{center}
		\includegraphics[width=8cm]{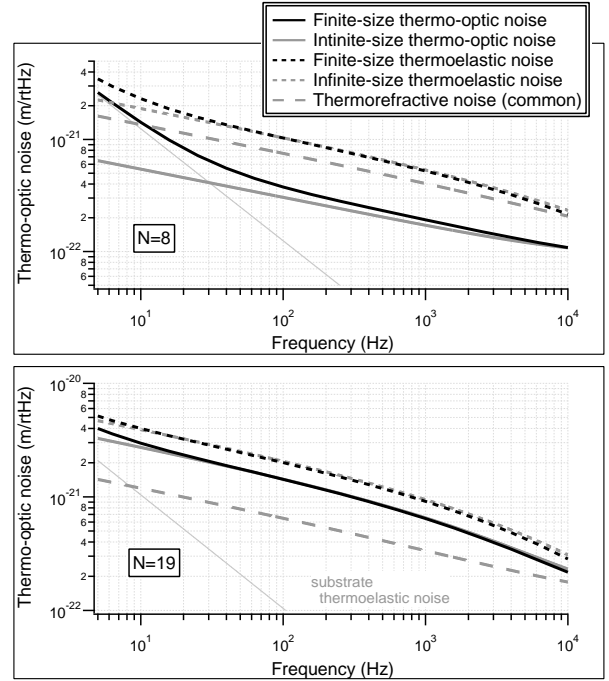}
	\caption{Thermo-optic-noise spectrum of each mirror in a Fabry-Perot arm cavity of Advanced LIGO detector~\cite{AdLIGO}. The mirror radius is 17~cm and the mirror thickness is 20~cm. The beam radius is 5.5~cm on the input test mass and 6.2~cm on the output test mass. The input test mass has 8 doublets, and the end test mass has 19 doublets of silica-tantala coatings.}
	\label{fig:TO}
	\end{center}
\end{figure}

Figure~\ref{fig:TO} shows the noise spectra of thermo-optic noise with a finite-size mirror and an infinite-size mirror. Here the averaged coatings are used. Compared with the result with an infinite-size mirror~\cite{Matt}, thermo-optic noise with a finite-size mirror is larger at low frequencies and smaller at high frequencies according to the difference in thermoelastic noise, while thermorefractive noise is same.

It is worth noting that there is a small, conceptual difference in the treatment of the two kinds of heat source for thermo-optic noise between this paper and Ref.~\cite{Matt}. Rigorously speaking, it is imaginary force that should be injected to calculate thermoelastic noise and it is imaginary entropy to calculate thermorefractive noise. In Ref.~\cite{Matt} it may seem like the entropy is injected for both purposes, and in our paper it may seem like the force is injected for both purposes. In fact, which imaginary quantity is injected does not matter since the conversion from the temperature fluctuation to displacement noise on the surface can be simply done by multiplying some constants. This is true as far as the stress inside the coating is uniform, or in other words, all the light is regarded to be reflected at the surface of the first coating layer.

\section{Summary}

We demonstrated the calculation of Brownian thermal noise and thermo-optic noise (thermoelastic noise $+$ thermorefractive noise) of the coatings on a finite-size cylindrical mirror. We used a method developed by Bondu {\it et al} based on the fluctuation-dissipation theorem to calculate the elastic response of the mirror, and extended it with the coatings. Comparisons with the previous calculations of thermal noise with an infinite-size mirror and with the independent calculation with a thin plate validate our results with a finite-size mirror. We showed how Brownian thermal noise and thermoelastic noise increase if the mirror becomes thin. We treated all the heat sources in the coatings and the substrate coherently, and the results give us the most accurate estimate of coating thermal noise.

\section*{Acknowledgement}
We would like to appreciate Prof.~Yanbei Chen, Dr.~Peter Fritschel, Dr.~Gregg Harry, Dr.~Yasushi Mino, Dr.~Calum Torrie, Dr.~Phil Willems, and Dr.~Hiro Yamamoto for valuable discussions. K.S. is supported by Japan Society for the Promotion of Science (JSPS). The research is also in some part supported by the Alexander von Humboldt Foundation's Sofja Kovalevskaja Programme.

\appendix
\section{Thermal noise of a thin plate}\label{sec:app}

We calculate thermal noise of a thin plate based on the study by Yamamoto {\it et al}~\cite{Yamamoto2007}. The noise level can be calculated with the modal-expansion method easily as the contribution of higher order modes is extremely small in the case of a thin mirror. This method is quite independent from the method we have shown in the main body of this paper, and the agreement of the results validates the calculation with a finite-size mirror.

\subsection{Brownian thermal noise}
With the modal-expansion method, the noise spectrum of the thermal motion in the first mode is given by
\begin{eqnarray}
S_x(\Omega)=\frac{4k_\mathrm{B}T}{m_1\omega_1^2Q_\mathrm{eff}}\frac{1}{\Omega}\ ,\label{eq:thinSx}
\end{eqnarray}
where $m_1$ and $\omega_1$ are the effective mass and the resonant frequency of the first mode, respectively, and $Q_\mathrm{eff}$ is the effective Q value of the coatings, which can be derived from the intrinsic loss angle and a compensation factor to adjust the elastic energy in the coatings to the total elastic energy~\cite{Yamamoto}:
\begin{eqnarray}
\frac{1}{Q_\mathrm{eff}}=\frac{3Y_\mathrm{c} d}{Y_\mathrm{s} h}\phi_\mathrm{c}\ .\label{eq:thinQ}
\end{eqnarray}
In the case of the thin-mirror analysis, contributions from the higher order modes are negligible, so that we can just calculate the contribution from the first mode. The elastic equation is 
\begin{eqnarray}
-\frac{h^2 Y_\mathrm{s}}{12(1-\nu_\mathrm{s}^2)}\triangle^2\varpi_1(r)=-\rho_\mathrm{s}\omega_1^2\varpi_1(r)\ ,
\end{eqnarray}
with $\varpi_1(r)$ as the one-dimensional displacement of the first mode and $\rho_\mathrm{s}$ as the density of the substrate. The boundary conditions are~\cite{Selvadurai}
\begin{eqnarray}
&&\left.\frac{d}{dr}\left(\frac{d^2\varpi_1}{dr^2}+\frac{1}{r}\frac{d\varpi_1}{dr}\right)\right|_{r=a}=0\ ,\nonumber\\
&&\left.\frac{d^2\varpi_1}{dr^2}+\frac{\nu_\mathrm{s}}{r}\frac{d\varpi_1}{dr}\right|_{r=a}=0\ ,
\end{eqnarray}
which gives the resonant frequency as
\begin{eqnarray}
\omega_1=\frac{\alpha_1^2}{a^2}\sqrt{\frac{Y_\mathrm{s}h^2}{12\rho_\mathrm{s}(1-\nu_\mathrm{s}^2)}}\ ,\label{eq:thinomega}
\end{eqnarray}
where $\alpha_1=2.9493$ is a solution of 
\begin{eqnarray}
\frac{2(1-\nu_\mathrm{s})}{\alpha_1}J_1(\alpha_1)-J_0(\alpha_1)-\frac{J_1(\alpha_1)}{I_1(\alpha_1)}I_0(\alpha_1)=0\ ,
\end{eqnarray}
with $J_n(\alpha_1)$ and $I_n(\alpha_1)$ as the Bessel function and the modified Bessel function of the first kind, and then $\varpi_1$ is given as
\begin{eqnarray}
\varpi_1=J_0\left(\alpha_1\frac{r}{a}\right)-\frac{J_1(\alpha_1)}{I_1(\alpha_1)}I_0\left(\alpha_1\frac{r}{a}\right)\ .
\end{eqnarray}
The effective mass is calculated from
\begin{eqnarray}
m_1=\frac{\displaystyle\int\rho_\mathrm{s}\left|\varpi_1\right|^2dV}{\displaystyle\left|\int\varpi_1 p(r)dS\right|^2}\ .\label{eq:thinm}
\end{eqnarray}
Plugging Eqs.~(\ref{eq:thinQ})(\ref{eq:thinomega})(\ref{eq:thinm}) into Eq.~(\ref{eq:thinSx}), we obtain the spectrum of Brownian thermal noise in the coatings of a thin plate.

\subsection{Thermoelastic noise}
Let us first derive thermoelastic noise caused by the heat in the coatings. Here we use the thin-coating approximation, and also we assume a mono-layer tantala coating. The solution of the heat equation can be resolved into the sum of functions that meet the boundary conditions that the heat flow should be zero at $z=0$ and $z=h$, then,
\begin{eqnarray}
\theta_\mathrm{s}=\sum_{n}{A_{n}\sqrt{\frac{2}{h}}\cos{\left(\frac{n\pi z}{h}\right)}}\ .
\end{eqnarray}
Plugging this into the heat equation, we get
\begin{eqnarray}
A_{n}&=&\frac{-i\Omega}{i\Omega+\kappa_\mathrm{s}(n\pi/h)^2}\frac{\alpha_\mathrm{eff}Y_\mathrm{s}Td}{C_\mathrm{s}(1-2\nu_\mathrm{s})}\Theta\times\sqrt{\frac{2}{h}}\ ,\nonumber\\
\alpha_\mathrm{eff}&=&\alpha_\mathrm{c}\frac{Y_\mathrm{c}(1-\nu_\mathrm{s})}{Y_\mathrm{s}(1-\nu_\mathrm{c})}-\alpha_\mathrm{s}\frac{C_\mathrm{c}}{C_\mathrm{s}}\ ,\\
\Theta&=&-\frac{1-2\nu_\mathrm{s}}{1-\nu_\mathrm{s}}\frac{h}{2}\left(\frac{d^2\varpi_1}{dr^2}+\frac{1}{r}\frac{d\varpi_1}{dr}\right)\ .\nonumber
\end{eqnarray}
The dissipation power is then given as
\begin{eqnarray}
W=\int\frac{\kappa_\mathrm{s} C_\mathrm{s}}{2T}\left|\frac{\partial\theta_\mathrm{s}}{\partial z}\right|^2dV\ ,
\end{eqnarray}
and the total energy is
\begin{eqnarray}
&&\hspace{-0.5cm}E_1^\mathrm{tot}=\frac{Y_\mathrm{s} h^3}{24(1+\nu_\mathrm{s})(1-\nu_\mathrm{s})}\nonumber\\
&&\hspace{0.5cm}\times\left[
\int\!\!\left| \frac{d^2\varpi_1}{dr^2}+\frac{1}{r}\frac{d\varpi_1}{dr} \right|^2\!rdr\right.\nonumber\\
&&\hspace{1.0cm}\left.-2(1-\nu_\mathrm{s})\int\!\left| \frac{d^2\varpi_1}{dr^2}\frac{1}{r}\frac{d\varpi_1}{dr} \right|\!rdr
\right]\ .
\end{eqnarray}
As the lost energy in one period is the total energy multiplied by $2\pi\phi$, the loss angle of the first mode caused by the heat source in the coatings is given as
\begin{eqnarray}
\phi_1^\mathrm{coa}&=&\frac{W}{\omega_1 E_1^\mathrm{tot}}\nonumber\\
&=&\frac{Y_\mathrm{s}\alpha_\mathrm{eff}^2T}{C_\mathrm{s}}\frac{1+\nu_\mathrm{s}}{1-\nu_\mathrm{s}}\frac{6d^2}{h^2}B_1\sum_{n}{\frac{\Omega\tau(n)^2}{(\Omega\tau)^2+(n)^4}}\ ,\label{eq:lossthin}
\end{eqnarray}
where $\tau=h^2/(\kappa_\mathrm{s}\pi^2)$ and
\begin{eqnarray}
&&\hspace{-0.5cm}B_1=\int\!\!\left| \frac{d^2\varpi_1}{dr^2}+\frac{1}{r}\frac{d\varpi_1}{dr} \right|^2\!rdr\nonumber\\
&&\hspace{0.5cm}\times\left[
\int\!\!\left| \frac{d^2\varpi_1}{dr^2}+\frac{1}{r}\frac{d\varpi_1}{dr} \right|^2\!rdr\right.\nonumber\\
&&\hspace{1.0cm}\left.-2(1-\nu_\mathrm{s})\int\!\left| \frac{d^2\varpi_1}{dr^2}\frac{1}{r}\frac{d\varpi_1}{dr} \right|\!rdr
\right]^{-1}\ ,
\end{eqnarray}
which is numerically calculated to be $1.47232$. Substituting $1/Q_\mathrm{eff}$ in Eq.~(\ref{eq:thinQ}) to $\phi_1^\mathrm{coa}$, we obtain coating thermoelastic noise. Note that the thin-mirror approximation let some errors in the result with a thin plate at high frequencies due to the thin-coating approximation; for example at frequencies higher than $\sim1$~kHz with $a=17~\mathrm{cm},\ h=5~\mathrm{cm},\ w_0=6.2~\mathrm{cm}$, and $N=19$. It is not a problem in Fig.~\ref{fig:TE} as the number of coatings is only 3.

For thermoelastic noise caused by the heat source in the substrate, the loss angle of the first mode is given as
\begin{eqnarray}
\phi_1^\mathrm{sub}&=&\frac{Y_\mathrm{s}\alpha_\mathrm{s}^2T}{C_\mathrm{s}}\frac{1+\nu_\mathrm{s}}{1-\nu_\mathrm{s}}B_1\frac{\Omega\tau}{1+(\Omega\tau)^2}\ ,\label{eq:lossthinsub}
\end{eqnarray}
the derivation of which is shown in Ref.~\cite{Blair}. Substituting $1/Q_\mathrm{eff}$ in Eq.~(\ref{eq:thinQ}) to $\phi_\mathrm{1}^\mathrm{sub}$, we obtain substrate thermoelastic noise.

\section{List of the parameters}\label{sec:appB}

\begin{itemize}
\item{Coating loss angle $\phi_\mathrm{j}$\\
$\mathrm{SiO_2}$\ :\ $1.0\times10^{-4}$,\ \ $\mathrm{Ta_2O_5}$\ :\ $4.0\times10^{-4}$}
\item{Thermal conductivity $k_\mathrm{j}$\\
$\mathrm{SiO_2}$\ :\ 1.38~W/m$\cdot$K,\ \ $\mathrm{Ta_2O_5}$\ :\ 33~W/m$\cdot$K}
\item{Thermal expansion $\alpha_\mathrm{j}$\\
$\mathrm{SiO_2}$\ :\ $5.1\times10^{-7}$/K,\ \ $\mathrm{Ta_2O_5}$\ :\ $3.6\times10^{-6}$/K}
\item{Specific heat per volume $C_\mathrm{j}$\\
$\mathrm{SiO_2}$\ :\ $1.64\times10^{6}$~J/K$\cdot$$\mathrm{m^3}$,\ \ $\mathrm{Ta_2O_5}$\ :\ $2.1\times10^{6}$~J/K$\cdot$$\mathrm{m^3}$}
\item{Thermal diffusivity $\kappa_\mathrm{j}\ (=k_\mathrm{j}/C_\mathrm{j})$}
\item{Young's modulus $Y_\mathrm{j}$\\
$\mathrm{SiO_2}$\ :\ $7.2\times10^{10}$~$\mathrm{N/m^2}$,\ \ $\mathrm{Ta_2O_5}$\ :\ $1.4\times10^{11}$~$\mathrm{N/m^2}$}
\item{Poisson ratio $\nu_\mathrm{j}$\\
$\mathrm{SiO_2}$\ :\ $0.17$,\ \ $\mathrm{Ta_2O_5}$\ :\ $0.23$}
\item{Refraction index $n_\mathrm{j}$\\
$\mathrm{SiO_2}$\ :\ 1.45,\ \ $\mathrm{Ta_2O_5}$\ :\ 2.06}
\item{Temperature dependence of the refraction index $\beta_\mathrm{j}$\\
$\mathrm{SiO_2}$\ :\ $8\times10^{-6}$/K,\ \ $\mathrm{Ta_2O_5}$\ :\ $14\times10^{-6}$/K}
\item{Density $\rho$\\
$\mathrm{SiO_2}$\ :\ 2200~$\mathrm{kg/m^3}$}
\item{Wavelength of light $\tilde{\lambda}$\\
1064~nm\ (Nd:YAG laser)}
\item{Temperature $T$\\
300~K}
\end{itemize}

\bibliographystyle{junsrt}

\begin{thebibliography}{99}
\bibitem{Note} Note that the definitions of $\kappa$ and $C$ are similar but different between references, for example between \cite{Liu} and \cite{Fejer}. Our paper follows the definition of \cite{Fejer}.
\end{thebibliography}
\pagestyle{headings}

\end{document}